# Evaluating research: from informed peer review to bibliometrics[1]

*Giovanni Abramo[a,b], Ciriaco Andrea D'Angelo[b]*

[a] National Research Council of Italy

[b] Laboratory for Studies of Research and Technology Transfer
School of Engineering, Department of Management
University of Rome "Tor Vergata"

**Abstract**

National research assessment exercises are becoming regular events in ever more countries. The present work contrasts the peer-review and bibliometrics approaches in the conduct of these exercises. The comparison is conducted in terms of the essential parameters of any measurement system: accuracy, robustness, validity, functionality, time and costs. Empirical evidence shows that for the natural and formal sciences, the bibliometric methodology is by far preferable to peer-review. Setting up national databases of publications by individual authors, derived from Web of Science or Scopus databases, would allow much better, cheaper and more frequent national research assessments.

**Keywords**

*Decision support systems; research assessment; peer review; bibliometrics; research productivity*

# 1. Introduction

National exercises for the evaluation of research work by universities and other public institutions are becoming regular events in ever more countries. In general, these exercises are aimed at informing selective funding allocations[2], stimulating better research performance, reducing information asymmetry between suppliers of new knowledge and their customers (students, companies, public administration, etc.), and last but not least, demonstrating that investment in research is effective and delivers public benefits.

Until recently, the conduct of these evaluation exercises has been founded on the so-called peer-review methodology, where research products submitted by institutions are evaluated by appointed panels of experts. In general, these assessments give the greatest weight to output quality. But recent developments in bibliometric indicators, particularly for measurement of publication quality, have lead many governments to introduce the more or less extensive use of these indicators in their next research assessments. The use of such measures is still limited to the natural and formal sciences[3], where publications in international journals and conference proceedings are the most accepted form for the diffusion of research outputs, and where the publications therefore represent a trustworthy proxy of research outputs (Moed, 2005). In the arts and humanities and most of the social sciences, bibliometric indicators are considered not yet sufficiently robust to inform peer-review. The penetration of bibliometrics in evaluation for the natural and formal sciences varies, as can be appreciated by examining the typologies of three of the upcoming assessment frameworks, from the most conservative to the most innovative: the Research Excellence Framework (REF) in the UK, the Quinquennial Research Evaluation (VQR) in Italy, and the Excellence in Research for Australia initiative (ERA). All three exercises are aimed at informing selective funding allocations. Submission activities for the ERA commenced in June 2010. Communication of VQR detailed guidance on submissions and assessment criteria is expected in 2010. The REF guidelines will be published during 2011; institutions will be invited to make submissions during 2013 and the assessment will take place during 2014. The ERA has an assessment period of six years, while other two have periods of five years. The considerations that follow, unless otherwise indicated, refer only to research evaluation for the natural and formal sciences.

The United Kingdom's REF is a typical example of a so called "informed peer-review" exercise, where the assessment outcomes will be a product of expert review informed by citation information and other quantitative indicators. The planners emphasize that judgments about the quality of individual outputs are not made solely on the basis of citation information: scoring of individual outputs must also always reflect expert judgment. Although the block grant is allocated at the level of higher education institutions, the REF will not assess the work of the whole research staff of each institution, but of selected research units that produce substantive bodies of work. The selection of the research units is to be done by the employing institution. The assessment will not entail all the research outputs of these units. It is proposed that each member submit a maximum of either three or four highest quality outputs[4].

---

[2] In some countries, such as the USA and Netherlands, the results of these exercises are not used to inform selective funding allocations.
[3] Mathematics and computer sciences, Physics, Chemistry, Earth sciences, Biological sciences, Medical sciences, Agriculture and veterinary sciences, Industrial and information engineering.

The Italian VQR may be considered a hybrid, a varying mix of pure peer-review, informed peer-review and the bibliometric approach, as selected by panels of experts for each of the 14 disciplines, which are in turn appointed by the Steering Committee for the Evaluation of Research (CIVR). To prepare judgments of quality, the panels of experts can then choose one or both of two methodologies for evaluating any particular output: i) analysis of citations; and/or ii) *peer-review* by external experts, selected by a collegial decision of the panel. Generally, the peer-review of an individual output is entrusted to a maximum of two experts, who give an anonymous opinion on the quality of the publication. Each university must submit two research outputs per researcher on staff, while other types of institutions must submit four. Each researcher is required to provide his/her institution with at least two (or four) outputs, indicating an ordering of their scientific importance. The institution is responsible for the final selection of the outputs to be submitted for the national evaluation. Based on the evaluation, block grants will then be allocated at the level of institutions, according to a ranking of the institutions derived from their performance in each discipline.

The Australian ERA assessment is conducted mainly through a bibliometric approach[5]. Single research outputs are evaluated by a citation index[6], relative to world and Australian benchmarks: no peer review is conducted in the natural and formal sciences. Because the entire research staff of the institutions must submit their full research product, indicators of research volume are also used to evaluate overall research performance.

Among the industrialized countries, the UK has the longest record of assessing research performance and linking funding to the outcomes of the assessment. Nevertheless, it shows more resistance, in wholly or partly replacing peer review with metrics, than newcomers in particular Australia. The members of the research evaluation and policy project (REPP) at Australian National University, have long sustained the need to develop robust quantitative measurement systems to evaluate research performance, which would obviate the need for a peer evaluation based on a large committee (Butler and McAllister, 2009).

Unfortunately, scientometrics research has focused much more on indicators than on the actual application of measurements. The aim of the present work is to contrast the peer-review and bibliometrics approaches in national research assessments. Accepting that there is no one infallible evaluation method, the position of the authors is that for the natural and formal sciences, the bibliometric methodology[7] is by far preferable to informed peer-review. We will support our position by comparing the two methodologies in terms of the essential parameters of any measurement system: accuracy, robustness, validity, functionality, time and costs (Table 1).

---

[4] Further information can be retrieved from "Second consultation on the assessment and funding of research" of September 2009, downloadable at http://www.hefce.ac.uk/pubs/hefce/2009/09_38/#exec. The launch of the REF is planned for 2012. Not all details are definitive and some may be subject to adjustment.

[5] The peer-review approach is used for the social sciences, arts and humanities. The list of the disciplines evaluated by bibliometrics only can be found at: http://www.arc.gov.au/era/key_docs10.htm

[6] If the research output submitted is published in a journal not indexed by Scopus, but is on the ERA journal list, it will be included in the 'ranked outlet' analysis but not used in 'citation analysis'.

[7] We refer to individual level research performance assessments through citation indicators. An example of such methodology is presented in Abramo and D'Angelo, 2011.

The factor that makes the bibliometric approach decisively preferable to peer review is not the better reliability of citation counts over peer judgment in assessing quality of individual outputs, which we see in the next section may be somewhat arguable, but rather other implications of the peer-review method. For obvious reasons of costs and time it would in fact be unthinkable to utilize peer-review to evaluate the entire output of a national research system. This results in undeniable penalties concerning the performance of peer-review as compared to bibliometric method, with respect to at least three of the above parameters. In turn, these shortcomings result from five major implications of limiting the evaluation to a subset of overall research output. First, it prevents any measure of productivity, the quintessential indicator of efficiency for any production system, and restricts evaluations to considering quality alone. Second, it seriously jeopardizes robustness of the measurement system, as seen in Section 3 of this report. Third, it hampers the validity of the measure, as will be demonstrated in Section 4. Fourth, it limits the functionality of the method, since it cannot be applied to single researchers or research groups, as shown in Section 5. Finally, with respect to the bibliometric approach, peer-review implies very high costs and times for execution, which limits its potential frequency of execution, as discussed in Section 6. The quantitative aspects of the comparative analysis for the two approaches refer mainly to the 2006 Italian Triennial Research Evaluation Framework (VTR), which was conducted using the peer-review method. The comparison is possible thanks to the database of the "Italian Observatory on Public Research" (ORP), derived by the authors from the Thomson Reuters Web of Science (WoS) databases. For the natural and formal sciences, the ORP permits comparative measurement of the research performance of every researcher in Italy. In these sciences, databases such as WoS by Thomson Reuters and Scopus by Elsevier offer quite thorough coverage of the journals most commonly used by scientists for disseminating their research outputs. The ORP and the bibliometric approach based on it will be presented in section 7. Conclusions will be drawn in section 8.

## 2. Accuracy

By accuracy of a research assessment system we mean the degree of closeness of performance indicators measurements to what is thought to be their actual value. In particular, the question posed in this section is whether the quality of a research output can be best evaluated by human judgment, or through the use of bibliometric indicators (citation and impact factor analyses), or by drawing on both (informed peer-review). It is frequently held that the basis of research evaluation is rightly that experts review the work of their colleagues. However, the exceptionally specialized nature of present-day research first makes it difficult to identify the most appropriate experts and then, given their acceptance to serve as reviewers, that they succeed in expressing fair judgments. The rapidity of scientific advances can also pose serious difficulties in contextualizing the quality of a research output produced a number of years previously. For example, a reviewer could very conceivably be called, in 2011, to evaluate a publication from 2004. Will he/she be capable of discounting all the subsequent intervening scientific advances in expressing a judgment? Peer evaluation is clearly susceptible to certain built-in distortions from subjectivity in assessments (Moxham and Anderson, 1992; Horrobin, 1990). These can occur at several levels, including the phase of using peer judgment to gauge output quality, but also in the earlier steps of selecting the experts that will carry out the assessments and, as will be shown later, selecting outputs to be submitted for evaluation. Subjective evaluations can be affected by real or potential conflicts of interest; from the inclination to give more positive evaluations to outputs from known and famous researchers than to those of the younger and less established researchers in the field; or from failure to recognize qualitative aspects of the product (a tendency that increases with increasing specialization and sophistication in the work). In addition, peer methodology does not have universal consistency, since mechanisms for assigning merit are established autonomously by the various evaluation panels and/or individual reviewers, thus exposing comparisons linked with this methodology to potential distortions. "Bias in peer review, whether intentional or inadvertent is widely recognized as a confounding factor in efforts to judge the quality of research" (Pendlebury, 2009).

On the other hand, bibliometric indicators can obviously not be applied to the entire range of research outputs, but only to publications and conference proceedings. In the sciences considered, the patent is also another important form of codification of research outputs, although less frequent. However, investigations have demonstrated a strong correlation between intensity of publication and intensity of patents (Adams and Griliches, 1998; and Lach and Shankerman, 2003). Another concern referring only to publications is that not all journals are indexed in the WoS or Scopus. Yet the most frequently heard objection to the use of citation counts in evaluating research is that citations do not always reflect quality. It has been shown though that negative citations may actually occur, but these are actually rare events and do not disrupt the analyses (Pendlebury, 2009). Another objection is that citation analysis is a less reliable proxy of quality for more recent works, especially in the formal sciences, where a longer lapse of time is necessary before citations mature. Even for more mature works, the phenomenon of "delayed recognition" (Garfield, 1980) may occur, and again this factor is often cited as a criticism of citation analyses. However, Glanzel (2008) showed that cases of delayed recognition are the exceptions to the rule. Finally, bibliometric indicators, such as the impact factor for journals and citation counts for authors, can be affected by certain forms of manipulation, which raises questions about their use in judging and ranking. However, the subjective judgment of reviewers can also be influenced by positive or negative attitudes towards one scientist or another, meaning that intentional bias can occur both through inflating citations by journals and authors and in altering the quality judgments given by reviewers.

The use of both methods, with the peer-reviewer having access to bibliometric indicators (hopefully appropriately standardized) concerning the publications to be evaluated, permits the reviewer to form an evaluation that emerges from comparison between his/her personal subjective judgment and the quantitative indicators. The pros and cons of the two approaches for evaluation of single scientific products are probably balanced, making it difficult to establish which would be preferable: the variables of context and the objectives of the evaluation could shift the weight in favour of one or the other. In fact, it is not an accident that many studies have demonstrated a positive correlation between peer quality esteem and frequency of citation (Abramo et al., 2009;

Aksnes and Taxt, 2004; Oppenheim and Norris, 2003; Rinia et al., 1998; Oppenheim, 1997).

A final comment is reserved for the process of selecting the reviewers and those who conduct this selection. This is a critical phase for the entire evaluation exercise, since the assurance of qualified reviewers is a necessary condition for the effectiveness of the entire process. The selection of reviewers requires total transparency, particularly as concerns the criteria applied. It seems likely, if not inevitable, that these criteria involve the use of quantitative indicators of scientific performance, as proxy of specific competence of the reviewer on the core topic of the product he/she has to assess. And if such indicators (publications and citations) are considered effective for identifying the best reviewers, they should be equally effective for the evaluation of single research output.

### 3. Robustness

In the specific case of a research performance measurement system we call it robust when it is able to provide a ranking which is not sensitive to the share of the research product evaluated. All peer-review research assessments limit comparative evaluation to a subset of the overall output of research organizations. The comparison of many institutions through the evaluation of only two research products, considered the best, for every researcher over a period of five years (VQR), or of 4 products from a subset of researchers at an institution (REF), could give results totally different than what would be obtained from considering a more ample subset or the entire research output. A check on robustness of the measurement system, through sensitivity analysis of institutions' research quality rankings to the share of research output evaluated, is thus rendered even more necessary. In the case of the past Italian VTR, the subset for evaluation was limited to a share of research outputs equal in number to 25% of the average count of full-time equivalent (FTE) researchers on staff at each university. For the three-year period evaluated, our investigations indicate that this number amounted to about 9% of the entire Italian university research output in the natural and formal sciences. Is it possible to base a comparison among research institutions on such a limited subset? The answer heard is that the evaluation fulfills the important role of assessing and providing incentives for excellence. But is it possible to identify, *a priori*, the ideal share of the total product? In comparing the exercises conducted in different nations it is clear that there is substantial variation in choosing the dimension for the subset of products to be submitted. One might ask if there is an optimum dimension, and whether this dimension should vary from one discipline to the next. In effect, because of the differing intensity of publication in the various disciplines, an evaluation based on a fixed number of products per researcher inevitably draws in unequal percentages of product from different disciplines. In the Italian VTR case, for example, the evaluation considered a range of between 4.6% of total output from Physics to 21.5% from Agriculture and veterinary sciences. Given such inconsistencies, we proceeded to check on robustness. To carry out such a test we first measured the correlation between the university rankings in the natural and formal sciences given under the VTR peer-review evaluation, based on publications submitted by universities, and the rankings obtained from bibliometric indicators of quality for the same publications. The correlation resulted significant in all of the eight disciplinary areas into which the natural and formal sciences are grouped in the Italian system, and strong in six of them (Abramo et al., 2009). Next we conducted a sensitivity analysis of the rankings, as measured by bibliometric indicators of quality, to variation in the share of product evaluated (Abramo et al., 2010). Shifting from a share corresponding to 25% of the FTE researchers (as in the VTR framework) to a consistent 9% share of total output in each discipline, which results in the same amount of output evaluated, was found to cause a notable variation in the quality ranking of universities. In Physics, for example, there is a change in ranking for 40 out of 50 universities, with a jump of up to 15 positions. In Biological sciences, the same change in the evaluation subset brings different rankings for 45 out of 53 universities, with a maximum shift of 22 positions. Or again taking Physics as an example, but considering eight different sizes of the output subset, from a minimum of 4.6% (as in the VTR framework) up to a maximum of 60% of the total output, only 8 universities out of 50 receive rankings that stay constant in the same decile (Table 2).

To identity a possible convergence of rankings for subsets of output over a certain size, we examined the variability of rankings with variation in the share of product submitted against a new benchmark: the evaluation of all universities' publications indexed in WoS. Analyzing the trends in the coefficient of correlation and the median variation in rankings of each scenario from the benchmark in Physics, there is a readily apparent convergence of rankings with increasing dimension of the subset (Figure 1). Also, beyond the "30% scenario", the increments in subset size achieve only marginal increases in correlation to the evaluation of all the publications. The same analysis was repeated for Biology: in this discipline, the stabilization effect of increasing subset size is less evident compared to that seen in Physics: the correlation with the benchmark seems to increase approximately in a linear manner over the whole spectrum of size scenarios considered (Figure 2) The conclusion is thus that the peer-review is intrinsically and inevitably lacking in robustness, and that even increasing the size of the subset of output evaluated, the variability of rankings is not reduced in an acceptable manner for all the disciplines. The empirical evidence unequivocally contradicts the founding assumption of the UK's next REF: "Experience of previous assessments demonstrates that assessing a sample of work of the highest quality is sufficient to provide a robust quality assessment in this context ..."[8].

If it is truly excellence that one wishes to assess and reward, then the best procedure would be to identify, from the overall output of all national institutions, the subset of products above a certain threshold of quality, to then observe their distribution among the institutions, standardize them with respect to the production factors of each institution, and formulate a final ranking list on the basis of the standardized values. The bibliometric method permits this.

**4. Validity**

In keeping with the memorable Albert Einstein's aphorism: "Not everything that can be counted counts, and not everything that counts can be counted", we define validity of a research assessment system its ability of the system to measure what counts. The fact that peer-review evaluation is necessarily limited to a subset of the entire research output also compromises its validity. The main objective of the national peer-review exercises is to evaluate the quality of research in organizations. However, in reality these exercises evaluate the quality of the products submitted which, because of the subjective process of their selection, are not necessarily the best of those produced by the institutions. For Italy's last VTR the number of research outputs submitted was set, per discipline for each institution, in function of the number of researchers belonging to that discipline. The next VTR, like the past Research Assessment Exercise (RAE) in the UK, will consider a set number of research products from every researcher. The UK's new REF actually restricts the evaluation, first to a subset of research staff from each institution and then to a subset of the research output of each of these individuals. The REF is thus exposed to the risk of further distortion, beyond that connected to the selection of the best products, linked to the possibility that the institutions will not necessarily identity their best researchers.

---

[8] From REF: "Second consultation on the assessment and funding of research" of September 2009, downloadable at http://www.hefce.ac.uk/pubs/hefce/2009/09_38/#exec

There is a potential that individuals or groups within an institution may sometimes exercise authority to somehow favor or block individuals or their products, instead of basing selection on intrinsic quality. The selection of a colleague or his/her research output could signify an unwelcome acceptance that the colleague (or output) is "better" than oneself (or one's own output). More fundamental, there are objective technical difficulties in comparing and selecting research products from different periods or subfields of research. Considering the past Italian VTR framework, one can imagine, as an example, the difficulty of a university faced with selection of a single best product from among three publications in the discipline of Medicine: one each from dermatology, cardiology and neuroscience. Or one could consider the difficulty of a single researcher who must choose two products from among all those he has produced in a five year period, and as is often the case, with these publications reflecting activity in various subfields of research. Abramo et al. (2009) carried out an investigation on the effectiveness of the selection process by universities in the past Italian VTR. Table 3 shows, for each discipline, the average percentages of publications that each university selected for the VTR for which the value of bibliometric quality indicator is lower than the median of the quality indicator distribution for all of the university's outputs in that discipline. Such average percentages show a range of variation from a minimum of 3.7% in biology to a maximum of 29.6% for agricultural and veterinary sciences. Other than this last discipline, notable figures also emerge for industrial and information engineering (26.5%) and mathematics and computer science (24.8%) as disciplines in which the selection process results as particularly ineffective. The data seen in the fourth column of Table 3 indicate that in 6 out of 8 disciplines there were actually universities that submitted all publications with a bibliometric quality indicator lower than the distribution median for the discipline.

The investigation confirms that the necessity of selecting a subset of products (or researchers) introduces elements of distortion that compromise the validity of the peer-review measurement system. In many cases the outputs evaluated do not represent the best products of the institutions. The inefficiency of the selection process means that the resulting rankings do not reflect the real quality of the institutions.

## 5. Functionality

We define functionality of a measurement system its ability to serve all the functions it is used for. National research evaluation exercises serve in the pursuit of a variety of policy objectives. First among these, in a number of nations, is the efficient allocation of resources. But this macroeconomic objective will not result from simply allocating resources to the best institutions: they in turn need to allocate resources to their best individual researchers and/or research groups. Peer-review types of national evaluations do not offer any assistance to universities for this concern, since they do not consistently penetrate to such precise and comparable levels of information. One might assume that research institutions would function as rational economic actors, allocating the resources they receive in function of merit in order to maximize their relative performance, and thus acquire more funding. However, such an expectation could be illusory. For the same motives as described earlier, during the phase of internal allocation of resources, the personal interests of individuals and groups exercising decisions could contrast with the collective interest of the institution. In Italian universities, for example, 28% of the research staff produces approximately 72% of the entire research output. The majority of researchers would thus not obtain direct advantage from an efficient allocation of resources and, presumably, would oppose such allocation. Even in cases where an institution intends to conduct selective funding allocation it may not necessarily have the technical means to compare the excellence of research from scientists working in different subfields. In fact, in order to compare the quality of research outputs from its own staff, the individual institution would have to resort to further peer-review, with costs and organizational efforts that would be difficult to sustain for the single institution. If such an institution would like to adopt bibliometric techniques to compare the production efficiency of researchers in different subfields it would have to carry out field-category standardizations in order to avoid the distortions related to the differing intensity of citations in different fields of research. The institution would need to first measure the quality of every output, standardizing with respect to a significant national or global benchmark, then compare the performance of each researcher (from the standardized values of the quality of his/her outputs) to those of colleagues in the same field of research. The process requires access to the raw data from bibliometric databases such as the WoS or Scopus and, as will be described, then demands a laborious task of "cleaning" the data and identifying the precise authorships. The process of comparative measurement of the performance of an institution's researchers, described in detail by Abramo and D'Angelo (2011), implies notable investments to acquire the raw bibliometric data and develop appropriate expertise for their treatment and successive elaboration. To conduct efficient internal selective funding allocations would require a duplication of processes at the level of single institutions and, if actually possible, would be very inefficient from a macroeconomic perspective.

National research assessments that measure and compare performance at the level of individual researchers and research groups would definitely be more functional not only for internal allocation of resources, but also for other policy objectives (stimulating individual performance, reducing information asymmetry, etc.). As we will see, the bibliometric method, unlike peer-review, has the full capacity to serve for this purpose.

## 6. Cost and Time Effectiveness

The direct costs of peer-review exercises are very high, varying with the number of products evaluated. For example, the UK's 2008 RAE, which evaluated four outputs per university researcher, cost 12 million pounds[9]. Again citing the example of the RAE, the indirect costs to the institutions evaluated, in terms of opportunity costs for the administrative and research staff time devoted, are estimated at five times the direct costs. The direct costs of the upcoming Italian VQR are estimated at 11 million euros. Meanwhile, the time necessary for implementation of peer-review exercises is long: two years or more for the entirety of the steps involved. All this means that these exercises typically occur over cycles of 5 to 6 years (the latest RAE actually covered an eight year period), which is slow and infrequent compared to what is necessary for efficient stimulation of improvement in research systems. Furthermore, researchers do not appreciate the time taken from their research activity to carry out the administrative practices and participate in "negotiations" concerning the selection of products to be submitted. The discontent of the researchers naturally becomes greater as they develop doubts concerning the value of the evaluation process or observe scarcity in the funds allocated based on its results. For their part, the institutions complain that the funding is allocated on the basis of scientific performance dating back to seven or eight years prior and demand greater frequency in the evaluation exercises.

ORP's direct costs, instead, are estimated at around 10% of peer-review exercises direct costs, and time required for execution is just few months.

## 7. ORP and the bibliometric approach

Even considering the limitations of bibliometrics cited above (especially the use of publications as a proxy of the entire research output and the use of citation counts as a proxy of the quality of publications), the superiority of bibliometrics over peer-review is evident for the natural and formal sciences, along the dimensions of:

- Robustness: bibliometrics allows evaluation of all, rather than a subset of overall output;
- Validity: it avoids any distortions that could occur during internal selection of products to be evaluated;
- Functionality: in providing evaluations for single scientists, then proceeding step by step to research groups, and ever larger aggregations, it permits each institution to allocate resources in an efficient manner;
- Cost and time effectiveness: it provides a dramatic saving on direct and indirect costs, and dramatically reduces time of execution.

Finally, bibliometrics is not limited to the evaluation of quality of research, but also permits the consideration of quantity.

However, the bibliometric method adopted in Australia still involves the submission of products selected by single institutions. This implies that each university must set up and update an archive of its scientific production, with data furnished by the researchers themselves, which implies both direct and indirect costs (opportunity costs for the researchers). A further problem is that basing research assessments on data provided by the actual research institutions engenders serious risks, linked to the inevitable errors in the data entry and in their subsequent submission to the national exercise. These problems are demonstrated by an earlier experience of this kind in Australia, the Composite Index. Audits conducted by KPMG found a high error rate in publication lists submitted by universities, especially at the outset of the application of this approach (34% in 1997). This error rate caused 97% of errors in the final scores and consequent funding allocations (Harman, 2000), although there has recently been a notable reduction in these rates[10].

---

[9] Research Excellence Framework, page 34, downloadable at www.hefce.ac.uk/pubs/hefce/2009/09_38/.

This context prompts an obvious question: if the bibliometric measures used for evaluating research outputs are derived from databases such as the WoS and Scopus, then why not use these databases for the direct extraction of the publications by individual researchers, thus avoiding all the consequences and processes of submission? The pertinent answer until now has been that deriving national bibliometric databases is a formidable task, because of difficulties involved in i) identifying and reconciling the varying ways in which authors of publications indicate the name of their "home" organization and ii) identifying the precise author of each publication, particularly because of homonyms among names (Aksnes, 2008) and variations in the way individual authors provide their names. This is why, until recently, remote bibliometric evaluations of research had been limited to single institutions, or in some cases to broader groups of institutions but dealing only with a limited number of disciplines. As recently as 2008, van Raan commented on the complete lack of national databases on the scientific production of individual researchers. Referring to a dataset of 18,000 WoS listings of publications by researchers in 10 Dutch universities, but only for those in the discipline of chemistry, he stated: "This material is quite unique. To our knowledge, no such compilations of very accurately verified publication sets on a large scale are used for statistical analysis …".

The authors have recently developed just such a national database, the ORP, derived from the WoS. Briefly, the ORP[11] lists all scientific publications produced since 2001 (about 272,000 articles and reviews, and 100,000 conference proceedings) by all public research organizations in Italy (approximately 350 in total). The ORP database succeeds in attributing all the publications to every academic author with an error of less than 5%. The underlying procedures for identifying and reconciling the author's institutional affiliation and the algorithms for author name disambiguation are described in D'Angelo et al., 2011. The bibliometric quality indicators are standardized with respect to the intensity of citations of the relevant subject category (which is not always the same as the author's discipline)[12]. This dataset is highly representative of the entire national research output for the natural and formal sciences (the ORP includes approximately 95% of the products presented to the VTR for these fields) and a few fields of social and economic sciences, i.e. for the output of 70% of the total research staff of Italian universities. Based on the ORP, an evaluation support system has also been developed, with the potential for producing rankings according to a number of performance indicators (productivity, productivity weighted for quality, productivity weighted by the number of co-authors, by the order of the author's co-listing, etc.). These indicators can be applied to measure the performance of each Italian university researcher[13] active in identified fields of research, limiting the distortions caused by the differing publication "fertility" and citation intensity that concern the various subject categories of research. Since Italian university researchers are classified in Scientific Disciplinary Sectors (SDSs), of which there are 370 in all, and since these are gathered into 14 "University Disciplinary Areas" (UDAs), it is possible to aggregate the data for single researchers and so proceed to measure the ranking of entire SDSs. Then, weighting their constituent SDSs for size, assessment can also proceed to broader rankings of UDAs and entire universities. Furthermore, turning individual performance ratings into percentile or standardized ranks, it is possible to compare units within a single university (research groups, departments, institutes, schools), considering their composition of researchers of from different disciplines. Such measures can even be carried out with standardization for academic rank of the personnel (full, associate and assistant professors), to compare the results achieved by the researchers at these different academic levels. Since the ORP starts from listings of the production of single researchers, it permits comparative evaluations of organizations based on the full complement of research staff, or restricted to the personnel actually active in publication, or limited to the top scientists alone (evaluation of excellence). From the analysis of co-authorship it is also possible to evaluate levels of activity in international research cooperation and the intensity of public-private collaboration.

---

[10] Audits of more recent exercises reported much lower levels of error, with the latest rate being under 10%, probably due to Australian universities learning how to better collect data on publications.
[11] More details on the ORP can be found in Abramo et al., 2008.
[12] A pertinent example is that of J. Hirsch, father of the bibliometric indicator by the same name, who is a physicist that publishes both in physics and in scientometrics categories.

The decision support system based on the ORP is absolutely non-invasive, since it does not require any input by the research institutions under observation. This offers savings in indirect costs and time for execution of evaluations. The lower costs would permit greater frequency of evaluations: on the order of months, rather than years.

## 8. Conclusions

The effectiveness of research evaluation systems and the indicators used can only be expressed relative to their intended objectives. National research evaluation exercises are ever more oriented towards bibliometric types of methodologies and indicators. The advantage of bibliometry with respect to classic peer review rests not so much in greater effectiveness at evaluating single research outputs, as in the possibility of evaluating all the publications indexed in databases such as the WoS and Scopus, which are highly representative of the entire research output in the natural and formal sciences. This certainly does not rend bibliometric evaluation perfect, but certainly makes it better than peer-review in terms of robustness, validity, functionality, costs and time of execution. Australia's 2010 ERA initiative uses the bibliometric method alone for comparative evaluation in the natural and formal sciences and the peer-review method for the social sciences, arts and humanities. Compromise methods, such as informed peer review, in which the reviewer can also draw on bibliometric indicators in forming a judgment, do not, in the opinion of the authors, offer advantages that justify the additional costs: indicators will not assist in composing human judgments, at the maximum permitting a confirmation or refutation. Assuming that individual judgments might be better than informed peer-review, all the typical limits of peer review would still remain, deriving from the small size of the subset of products evaluated. The ERA framework involves the submission of research products by institutions, which draw on their internal archives of research outputs, involving high opportunity costs for the researchers that enter the data and duplications and inevitable errors in transcription that produce distorting effects for the final rankings of the universities.

---

[13] A number of Italian universities (e.g., the universities of Rome Tor Vergata, Milan, Pavia, Cagliari and Udine) have already used the ORP system for comparative evaluation of research.

Scientometrics literature abounds with works concerning the ideal indicators of scientific excellence. Every letter of the alphabet has been applied to naming them and their variants. Even though the choice of the most appropriate indicators is fundamental for any measurement, an evaluation also necessitates a broadly applicable benchmark to meet the demands of both standardization and comparison. Unfortunately, scientometrists have not applied equal energy to developing wide-scale measurement systems based on their proposed indicators. Attention has been much more focused on the indicators than on the methodology of applying the measurement. Italy's ORP results as the only national database derived from international bibliometric databases, such as the WoS or Scopus, that attributes publications to their precise authors within acceptable error (5%). For evaluation exercises at the aggregate level, such as among large-sized organizational units (institutions, departments, disciplines, etc.), this level of false negative and false positive authorships, being distributed with sufficient uniformity among the subjects of the evaluation, causes distortions in ranking that can still be readily accepted. However for comparative evaluation at the individual level, such errors could result in cases of strong distortion, meaning that it would be advisable to have a further check of the authorship by the authors themselves and/or the judgment of the unit head. In the conception of the authors, the ORP is presented not as an evaluation system in itself, but rather as a support to evaluation, analogous to the diagnostic imaging tools used in medicine. The results of an x-ray examination might be sufficient for a specialist to prepare a diagnosis, but there could be a need for further examination. For evaluation at the micro level of small-sized research units, further checking is always desirable. For national exercises of evaluation, databases such as the ORP could avoid the step of institutions submitting publications, resulting in less error in subsequent measurements and rankings, and notable advantages in costs and times of execution, which would permit greater frequency in running such exercises.

It is now possible, at least in Italy, for a national research evaluation to use bibliometric methods for the natural and formal sciences, based on data such as those found in the ORP, rather than on submissions from institutions, and to apply peer-review methods for other disciplines. Development of similar databases in other nations would offer the useful possibility of international comparisons. Evaluation systems that contribute to comparative measurement of performance by single researchers and research groups also permit greater efficiency in the internal incentive systems of research organizations, in recruitment, in informing selections in the general market for knowledge, in analysis of research strengths and weaknesses at the regional level, and so in formulating research and industrial policy at both national and regional level.

Given the current development of bibliometric indicators relative to the development of actual methodologies for applying their use, the hope is that scientometrists in other nations will now join in concentrating on the resolution of problems involved in large-scale measurement and on the construction of national databases that permit the implementation of such measures. So far, peer-review still remains the more appropriate methodology to evaluate the quality of research in disciplines other than the natural and formal sciences.

| Parameter | Description |
| --- | --- |
| Accuracy | The degree of closeness of performance indicators measurements to their true value |
| Robustness | The ability of the system to provide a ranking which is not sensitive to the share of the research product evaluated |
| Validity | The ability of the system to measure what counts |
| Functionality | The ability of the measurement system to serve all the functions it is used for |
| Time | The time needed to carry out the measurement |
| Costs | The direct and indirect costs of measuring |

*Table 1: Parameters to evaluate research performance measurement systems*

| University | \multicolumn{10}{c}{Ranking decile} |
| | 1 | 2 | 3 | 4 | 5 | 6 | 7 | 8 | 9 | 10 |
|---|---|---|---|---|---|---|---|---|---|---|
| Scuola Normale Superiore in Pisa | 100 | | | | | | | | | |
| International School for Advanced Studies of Trieste | 100 | | | | | | | | | |
| University of Venice "Ca' Foscari" | 75 | 13 | 13 | | | | | | | |
| University of "Roma Tre" | 100 | | | | | | | | | |
| University of Verona | 50 | 13 | | | 13 | | 13 | | | 13 |
| University of Varese "Insubria" | 38 | 63 | | | | | | | | |
| University of Trent | 25 | 75 | | | | | | | | |
| Polytechnic University of Turin | | 88 | 13 | | | | | | | |
| University of Rome "La Sapienza" | | 88 | 13 | | | | | | | |
| Polytechnic University of Milan | | 25 | 63 | 13 | | | | | | |
| Second University of Naples | | 25 | 25 | 13 | | 13 | 13 | 13 | | |
| University of Camerino | | 38 | 38 | | 13 | | | 13 | | |
| Polytechnic University of Bari | | 38 | 25 | 25 | 13 | | | | | |
| University of Pisa | | | 88 | 13 | | | | | | |
| University of Milan "Bicocca" | | | 63 | 13 | 13 | 13 | | | | |
| University of Ferrara | | 13 | 38 | 25 | 13 | 13 | | | | |
| University of Florence | | | | 100 | | | | | | |
| University of Trieste | | 25 | 63 | 13 | | | | | | |
| University of Rome "Tor Vergata" | | 13 | 38 | 25 | 13 | 13 | | | | |
| University of Salerno | | 13 | 25 | 13 | 50 | | | | | |
| University of Padua | | | 38 | 63 | | | | | | |
| University of Perugia | | | 50 | 50 | | | | | | |
| Polytechnic University of Ancona | | | 25 | 13 | 25 | 13 | | 25 | | |
| University of Modena and Reggio Emilia | | | | 13 | 50 | 38 | | | | |
| University of Milan | | | | | 50 | | 38 | 13 | | |
| University of Bologna | | | | | 13 | 38 | 50 | | | |
| University of Turin | | | | | 50 | 25 | 25 | | | |
| University of Pavia | | | | | 13 | 88 | | | | |
| University of Udine | | | | | 13 | 38 | 25 | 25 | | |
| University of Catania | | | | | 38 | 25 | 38 | | | |
| University of L'Aquila | | | 13 | | | 38 | 50 | | | |
| University of Basilicata | | | | | 38 | 13 | 25 | 25 | | |
| University of Genoa | | | | | 13 | 63 | 25 | | | |
| University of Eastern Piedmont "A. Avogadro" | | | | | 13 | | 50 | 38 | | |
| University of Lecce "Salento" | | | | 13 | | 25 | 50 | 13 | | |
| University of Brescia | | 13 | 25 | | 13 | | | 50 | | |
| Sacred Heart Catholic University | | | | 25 | | 13 | 38 | 25 | | |
| University of Urbino "Carlo Bo" | | 13 | 25 | | | 13 | | 50 | | |
| University of Naples "Federico II" | | | | | | | 100 | | | |
| University of Cagliari | | | | 25 | | 13 | | 50 | 13 | |
| University of Bari | | | | | 13 | | 13 | 13 | 63 | |
| University of Parma | | | | | | | | | 100 | |
| University of Calabria | | | | | | | | | 63 | 38 |
| University of Messina | | | | | | | | | 63 | 38 |
| University of Palermo | | | | | | | | | 88 | 13 |
| University of Chieti "Gabriele D'Annunzio" | | | | | | 13 | | 25 | 63 | |
| University of Siena | | | | | | | | | 13 | 88 |
| University of Sassari | | | | | 13 | | | | 38 | 50 |
| University of Benevento "Sannio" | | | | | | | | | | 100 |
| University of Naples "Parthenope" | | | | | | | | | | 100 |

*Table 2: Frequency matrix of Italian university ranking, in deciles, under various scenarios of publication share in Physics. Bibliometric simulation based on average impact ratings of 2001-2003 WoS data*

| Discipline | Average | Median | Max | Variation coefficient |
| --- | --- | --- | --- | --- |
| Agricultural and veterinary science | 29.6% | 26.3% | 100% | 0.912 |
| Industrial and information engineering | 26.5% | 26.0% | 100% | 0.868 |
| Mathematics and computer science | 24.8% | 24.0% | 100% | 0.897 |
| Earth science | 17.4% | 14.3% | 100% | 1.179 |
| Physics | 8.5% | 0% | 100% | 1.939 |
| Chemistry | 5.0% | 0% | 100% | 3.312 |
| Medicine | 3.8% | 1.2% | 33.3% | 1.895 |
| Biology | 3.7% | 0% | 35.3% | 2.105 |

*Table 3: Statistics for percentages of outputs which each university selected for the VTR, where the value of the bibliometric quality index is lower than the median of the quality index distribution of the total outputs by the university.*

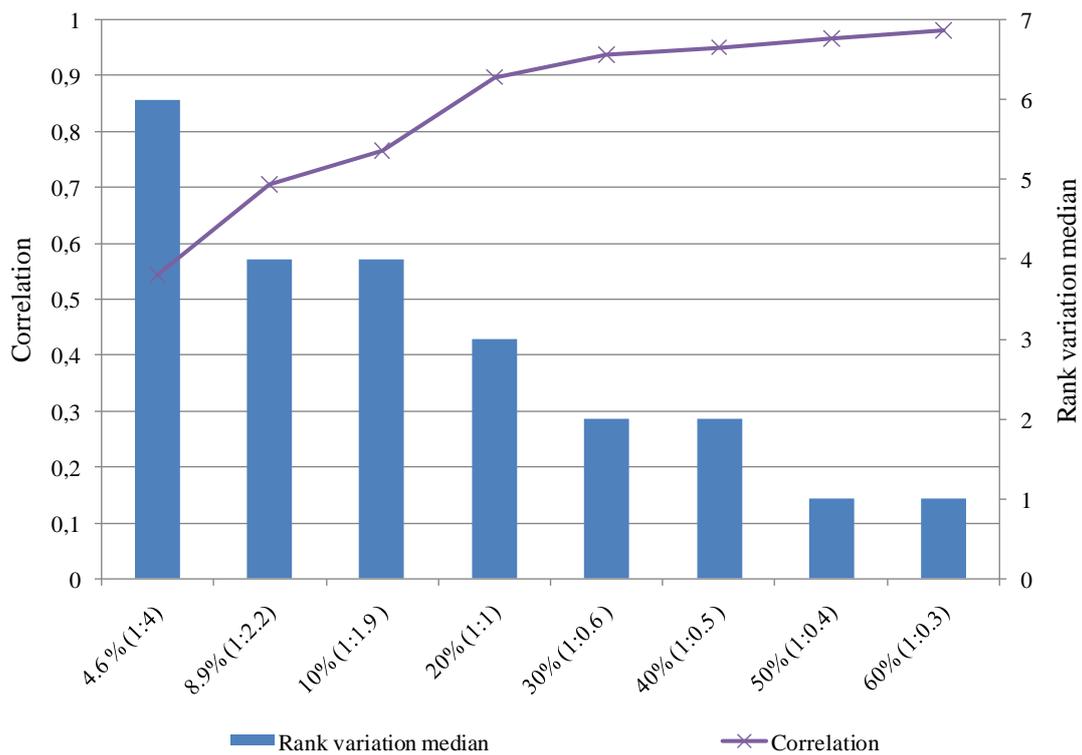

*Figure 1: Trends in the coefficient of correlation and median for variation of university research quality rank under various scenarios for product share, for the Physics UDA, with the benchmark being the evaluation of all products.*

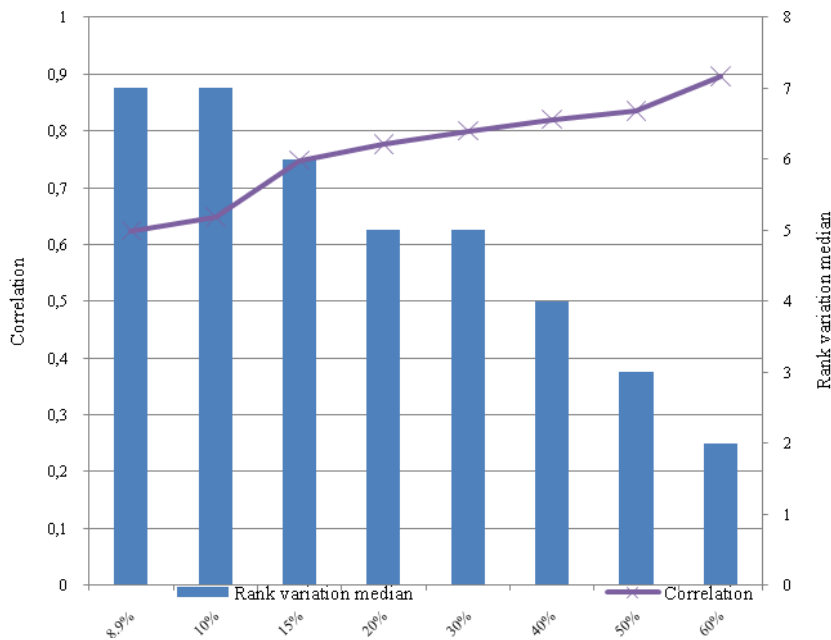

*Figure 2: Trends in the coefficient of correlation and median for variation of university research quality rank under various scenarios for product share, for Biology, with the benchmark being the evaluation of all products.*